\documentclass[10pt,a4paper,twoside]{aa}
\usepackage{aalongtable}
\usepackage[tight,scriptsize]{subfigure}
\usepackage{natbib}
\usepackage{graphicx}
\usepackage{amssymb}
\newcommand{\II}{\scriptsize{II}\normalsize}
\begin{document}
\title{Mg \normalsize{II} \Large{ h + k emission lines as stellar
    activity indicators of main sequence F-K stars} } 
\author{Andrea P. Buccino \and Pablo J. D. Mauas}
    \institute{ Instituto de Astronom\'\i a y F\'\i sica del Espacio
    (CONICET), C.C. 67 Sucursal 28, 1428-Buenos Aires Argentina}
    \offprints{A. P. Buccino, \email{abuccino@iafe.uba.ar}}
    \date{Accepeted: March 27$\textrm{th}$, 2008. } 
    \abstract{The largest dataset of stellar
    activity measurements available at present is the one obtained at
    the Mount Wilson Observatory, where high-precision Ca
    \scriptsize{II} \normalsize H+K fluxes have been measured from
    1966 for about 2200 stars. Since the Mg \II\- h and k
    lines at $\lambda$2800 \AA\- are formed in a similar way to the Ca
    \scriptsize{II} \normalsize H+K emission lines, they are
    also good indicators of chromospheric structure. The
    \emph{International Ultraviolet Explorer (IUE)} provides a large
    database of UV spectra in the band 1150-3350 \AA\-, from 1978 to
    1995, which can also be used to study stellar activity.}{The main
    purpose of this study is to use the IUE spectra in the
    analysis of magnetic activity of main sequence  F-K stars.  Combining IUE
    observations of Mg \II\- and optical spectroscopy of Ca \II, the
    registry of activity of stars can be extended in time.}{We
    retrieved all the high-resolution spectra of F, G, and K main
    sequence stars observed by IUE (i.e. 1623 spectra of 259 F to K
    dwarf stars). We obtained the continuum surface flux near the Mg
    \II\- h+k lines near $\lambda$2800 \AA\- and the Mg \II\- line-core surface flux from the IUE
    spectra.}{ We obtained a relation between the mean continuum flux
    near the Mg \II\- lines with the colour $B-V$ of the star.  For a
    set of 117 nearly simultaneous observations of Mg \II\- and Ca
    \II\- fluxes of 21 F5 to K3 main sequence stars, we obtained a
    colour dependent relation between the Mount Wilson Ca
    \II\- $S$-index and the
    Mg \II\- emission line-core flux. As an application of this calibration, we
    computed the Mount Wilson index for all the dF to dK stars which
    have high resolution IUE spectra. For some of the most frequently observed
    main sequence stars, we analysed the Mount Wilson index $S$ from
    the IUE spectra, together with the ones derived from visible
    spectra. We confirm the cyclic chromospheric activity of
    $\epsilon$ Eri (HD 22049)
    and $\beta$ Hydri (HD 2151), and we find a magnetic
    cycle in  $\alpha$ Cen B (HD
    128621)}{}

\titlerunning{Mg II lines}
\authorrunning{Buccino \& Mauas}
\maketitle
\section{Introduction}\label{intro}

Stellar magnetic activity causes non-thermal heating of the outer
atmospheres of cool stars. One of the principal diagnostics for solar
and stellar chromospheric activity is the emission in the Ca
\scriptsize{II} \normalsize H and K resonance lines (at 3968 and 3934
\AA). In particular, the largest dataset of activity measurements
available at present, comprising observations of over two thousands
stars, is the one obtained with the Mount Wilson HK
spectrophotometers, which since 1966  measure high precision Ca \scriptsize{II}
\normalsize H+K fluxes. As an indicator of
stellar activity, an index $S$ has been defined as the ratio between
the emission
line-core flux and the flux in the continuum nearby.

 Since the Mg \II\- h and k lines (at 2803 and 2796 \AA\-) are formed
in a similar way to the Ca \scriptsize{II} \normalsize H+K
lines, they are also good indicators of the heating and the
thermal structure of stellar atmospheres, especially from the high
photosphere to the upper part of the chromospheric
plateau. Furthermore, the Mg \scriptsize{II} \normalsize resonance
lines are more sensitive to weak chromospheric activity than the Ca
\scriptsize{II} \normalsize ones, because the adjacent near UV
continuum is significantly weaker in the Mg \scriptsize{II}
\normalsize continuum and the photospheric line wings are darker at
2800 \AA. 

In the solar case, the Mg \scriptsize{II} \normalsize core-to-wing
ratio  defined by \cite{1986JGR....91.8672H}, has become a valuable
index of  variability of the
chromospheric  radiation. For the last twenty years, solar activity has
been monitored from  space by many instruments (SBUV, SOLSTICE, SUSIM
and GOME). These
observations have provided  valuable data from which the Mg
\scriptsize{II}  \normalsize index can be
derived.

In order to connect stellar and solar observations,
\cite{1992A&A...256..185C} compared IUE Mg \II\- profiles with some
\emph{Skylab} spectra of solar regions. They showed that different Mg
\II\- emission levels observed in stars of similar spectral type are due
to differing fractions of their surfaces covered by magnetic regions.

In previous studies, it was found that the radiative fluxes of the Ca \scriptsize{II} \normalsize
and the Mg \scriptsize{II} \normalsize lines are highly correlated among  themselves (\citealt{1985A&A...147..265O},
\citealt{1987A&A...172..111S}, and \citealt{1991A&A...252..203R}). Using nearly simultaneous
observations, \cite{1992A&A...258..432S} derived a linear relationship
between the Mg \scriptsize{II} \normalsize h+k fluxes ($F_{Mg\,II}$),
measured using
IUE\footnote{\emph{International
Ultraviolet Explorer}} spectra, and the Mount Wilson Ca \scriptsize{II} \normalsize H+K fluxes
($F_{Ca\,II}$). 

The study of chromospheric variability requires at least a decade of data to
 reveal variations with timescales similar to the 11 yr solar cycle.
 IUE provides an extensive homogeneous  
 database of UV spectra covering the band 1150-3350 \AA\-, from 1978 to 1995.
The purpose of this study is to intercalibrate the Mg \II\- and Ca \II\-
 observations,  to combine IUE observations of Mg \II\- h+k and
 visible observations of Ca \II\- H+K and  to extend in time the registry of
 activity of solar-type and cooler stars.

In Section \S\ref{sec.uvcont}, we first analyse the UV continuum
adjacent to the  Mg
\II\- lines. In Section \S\ref{calindex}, we derive a relation to
 determine  the Mount Wilson index from the Mg \II\- line-core
flux. Finally, in Section \S\ref{appl} we study, for several main sequence stars, the
Mount Wilson  indices we  obtained
indirectly from IUE observation, the ones published by
\cite{1996AJ....111..439H} from CTIO spectra and the indices obtained
from CASLEO spectra.

\section{UV continuum near the Mg \scriptsize{II} \normalsize
  lines}\label{sec.uvcont} 

\begin{figure}[htb!]
\resizebox{\hsize}{!}{\includegraphics{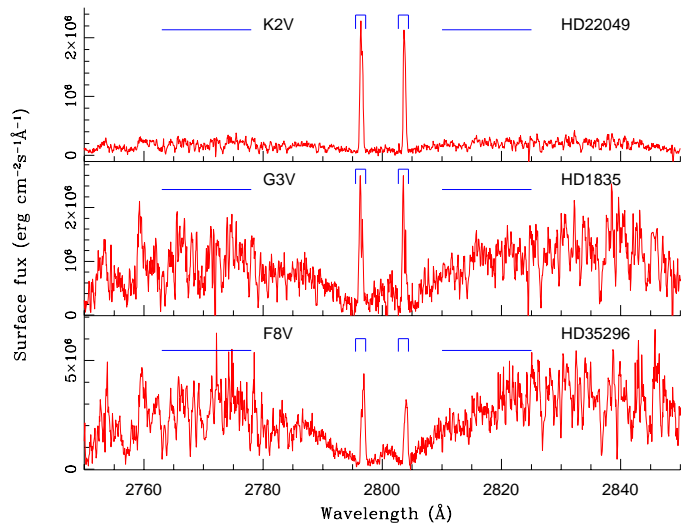}}
\caption{IUE spectra of three representative stars: HD 22049 (K2V), HD 1835 (G3V) and HD 35296 (F8V). 
  The   windows used to integrate the continuum flux (2763-2778  \AA\- and 2810-2825 \AA)\-  and the
  line-core  emission (Mg \scriptsize{II} \footnotesize k:
  2795.50-2797.20 \AA\-  and Mg \scriptsize{II} \footnotesize h:
  2802.68-2804.38 \AA)\- are marked.}\label{ind1}
\end{figure}

In Fig. \ref{ind1} we present some examples of the high
resolution IUE spectra of three F, G and K main sequence
stars. The spectra are available from the IUE public library (at
\textsf{http://ines.laeff.esa.es/cgi-ines/IUEdbsMY}), and have been
calibrated using the NEWSIPS (New Spectral Image Processing System)
algorithm \citep{1997IUENN..57....1G}.  The internal accuracy of the
high resolution calibration is around 4\%
(\citealt{2000A&AS..141..331C}).

 The measured  flux from the IUE spectra \emph{f} was transformed to  surface
flux  \emph{F} using the \cite{1982A&A...110...30O} relation
\begin{equation}
log\,(F/f)=0.35+0.4(m_V+BC)+4\,\,log\,T_{eff},\label{fluxcal}
\end{equation}
 where $m_V$ is the visual apparent magnitude that we obtained from
the \emph{Hipparcos and Tycho Catalogue}
(\citealt{1997A&A...323L..49P}, \citealt{1997A&A...323L..57H}), $BC$ is the
 bolometric  correction obtained from \cite{1996ApJ...469..355F} and
 $T_{eff}$  is
 the  effective temperature,  from \cite{1981ARA&A..19..295B}.

To analyse the UV continuum surface flux near the Mg \II\-
lines, we obtained all the high resolution spectra of F, G and K main
sequence stars observed by IUE (i.e. 1623 spectra of  259 F to K
dwarf stars), and we integrated the flux  in two  windows 15 \AA\- wide centred at
2817.50 \AA\- and 2770.50 \AA\, ($F_{cont_{MgII}}$, Fig. \ref{ind1}).  These windows are
as wide as possible to obtain the best signal-to-noise relation,
without including lines of chromospheric origin.

\begin{figure}[htb!]
\resizebox{\hsize}{!}{\includegraphics{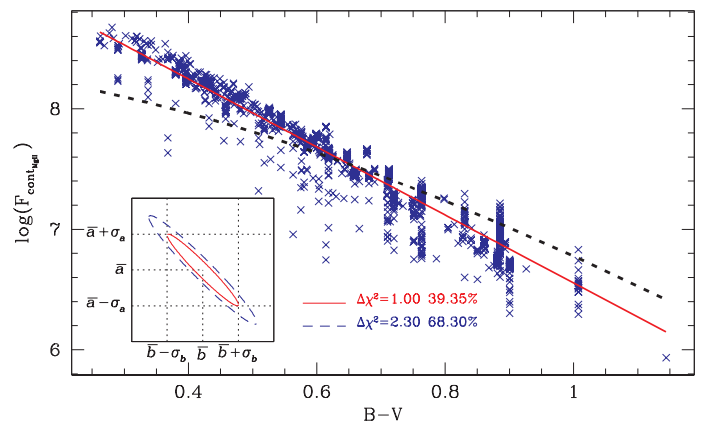}}
\caption{log ($F_{cont_{Mg II}}$) vs. $B-V$ for 1623 high
  resolution IUE  spectra of 259 stars. The solid line shows the
  best linear interpolation and the dashed line
  is the relation obtained by \cite{1984A&A...130..353R}
  for the visible continuum surface flux. In the inset, we show
  the $\chi^2$ contours of the 39.3\% (full line) and 68.3\% (dotted line)
  confidence levels for the two fitted parameters, assuming that they
  present normal distributions with  mean values and  standard deviations:
  $\overline{a}=-2.823$, $\sigma_a=0.018$ and $\overline{b}=9.376$, $\sigma_b=0.013$. }\label{cont}
\end{figure}

   In Fig.  \ref{cont},
the \mbox{log ($F_{cont_{Mg II}}$)} derived from each IUE spectrum is
plotted against the colour \mbox{$B-V$}, from the Hipparcos and Tycho Catalogues. In our set of
observations, the measured continuum flux presents a mean intrinsic
variation of 30\% for some stars; we will explore the source of this
dispersion at the end of this section. For simplicity, we do not include the error bars
in the figure. On the other hand, we neglect
the errors in $B-V$ since they are very small. 

For the data plotted in
 Fig.  \ref{cont}, we found  a
 good linear regression \mbox{$\log(F_{contMgII})=b(B-V)+a$}, with a correlation
 coefficient R=0.97. 
The best fit is given by
\begin{equation}
\centering 
\log(F_{contMgII})=(-2.823 \pm 0.018)(B-V)  +(9.376 \pm 0.013).
\label{eq.logont}
\end{equation}

Therefore,
 we obtain an
 exponential
 relation  between an average
  UV continuum surface flux and
 the colour \mbox{$B-V$}. The mean value $\langle F_{{cont}_{Mg
 II}}\rangle$ and the standard deviation
 $\sigma_{ F_{{cont}_{Mg II}}}$ are:
\vspace{-1cm}
{\setlength\arraycolsep{4pt}
\flushright
\begin{eqnarray}
\langle F_{{cont}_{Mg II}}\rangle=2.38\times10^9\times
10^{-2.82(B-V)}\,\textrm{erg}\,\textrm{cm}^{-2}\,\textrm{s}^{-1},\,\,\label{regre1}\\
\nonumber\\
\sigma^2_{ F_{{cont}_{Mg II}}}=\langle
F_{{cont}_{MgII}}\rangle^2\,P_2(B-V),\,\,\,\,\,\,\label{regre2}\\
\nonumber\\
\textrm{with}\,\,\,
P_2(B-V)=[9.5+1.3(B-V)+1.5(B-V)^2]\times 10^{-3}.\nonumber
\end{eqnarray}
}
The regression in Eq. \ref{regre1} is significant at
nearly a 70\% confidence level, which tests satisfactorily the linear
fit in Fig. \ref{cont}. 

In contrast, \cite{1989ApJ...341.1035S} assumed that the Mg
 \scriptsize{II} \normalsize continuum surface 
 flux  has the same dependence on colour than the one  obtained by
 \cite{1984A&A...130..353R}  for the Ca \scriptsize{II}\normalsize\-
 continuum. Rutten found that the
 logarithm of the visible  continuum
 surface flux  was proportional to a third
 order  polynomial on \mbox{$B-V$}.  In Fig. \ref{cont} we represent
 this  relation
 with a dashed curve. It can be seen that the linear relation we
 obtained fits the observations better.

\begin{figure}[htb!]
\resizebox{\hsize}{!}{\includegraphics{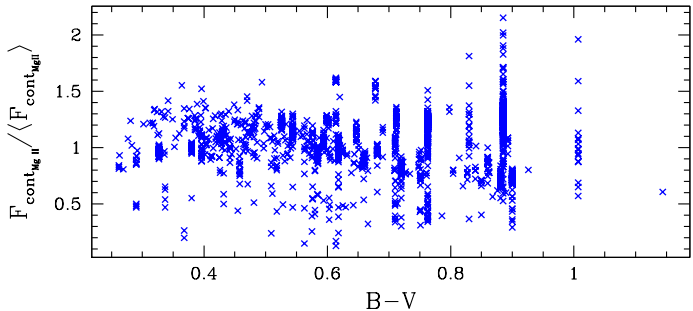}}
\caption{The continuum surface flux $F_{cont_{MgII}}$ normalized to
 the averaged surface UV continuum $\langle F_{cont_{MgII}}\rangle$
 vs. $B-V$ for the 1623 high resolution IUE
 observations.}\label{resnorm}
\end{figure}

To analyse whether the spread in the data in Fig. \ref{cont} could be
 due to a remaining colour-dependent component in the continuum flux,
 we plot in Fig. \ref{resnorm} the ratio of the UV emission
 $F_{cont_{Mg II}}$ to the value given by Eq. \ref{regre1}. The mean
 value of this ratio is 1.04 and the residuals present a flat
 distribution vs. $B-V$. We applied a Wilcoxon
 two-sample test \citep{Frod79} to the data plotted in Fig. \ref{resnorm} and we
 obtained that the fluctuations of the ratio $F_{cont_{MgII}}$/$\langle
 F_{cont_{MgII}}\rangle$ are within the statistical error with a
 confidence level of 90\%. Therefore, we conclude that the spread in
 the data is not associated with a colour dependent component.
\begin{figure}[htb!]
\resizebox{\hsize}{!}{\includegraphics{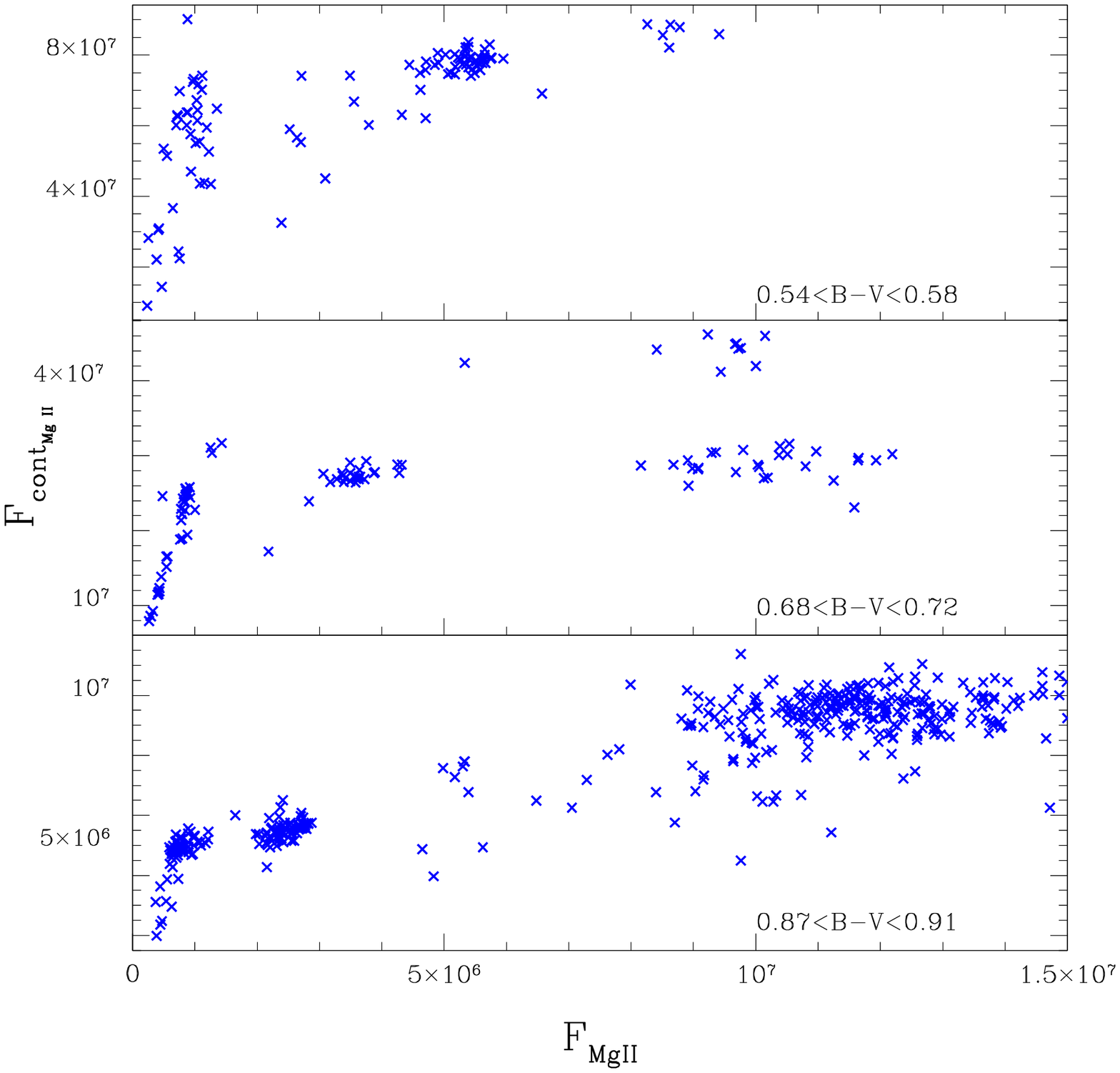}}
\caption{$F_{cont_{Mg II}}$ vs. $F_{MgII}$, for
  a set of main sequence stars with \mbox{0.54$<$$B-V$$<$0.58} (\emph{top}), \mbox{0.68$<$$B-V$$<$0.72} (\emph{middle}) and  \mbox{0.87$<$ $B-V$$<$0.91} (\emph{bottom}).}\label{cont_act}
\end{figure}

 On the other hand, since the Mg \II\- continuum is originated in the
upper photosphere and the lower chromosphere of the star, it can be
sensitive to activity.  In fact, since the vertical spread for a specific
colour $B-V$ in Fig. \ref{resnorm} corresponds, in most cases, to
different observations of  a single
star, it is probably due to different levels of 
chromospheric activity.  

 To check this, in Fig. \ref{cont_act} we plot  the continuum surface flux
$F_{cont_{Mg II}}$ vs.  the Mg \scriptsize{II} \normalsize line-core
emission $F_{MgII}$ for three different colour bins. To integrate the
Mg \II\-\ line-core fluxes,  we found that the best integration
window in the lines for high resolution IUE spectra  are two 1.70 \AA\- wide passbands centred at 2803.53 \AA\- and
2796.35 \AA\- (see Fig. \ref{ind1}). The position and the
width of the integration windows were chosen 
to guarantee that the contribution of the integrated flux is merely
chromospheric, beyond the basal contribution.

The  Ca \scriptsize{II}\normalsize\- continuum flux, used in the Mount
Wilson index definition, is associated with photospheric emission and
 chromospheric activity is insignificant in this wavelength range. In
contrast, in Fig. \ref{cont_act} we note that, independently from the colour, the $F_{cont_{Mg II}}$
increases with the Mg \scriptsize{II} \normalsize line-core flux and,
therefore, the continuum flux depends to some extent on the activity
level. For this reason, we did not attempt to build an activity index as the
ratio in the fluxes in the Mg \II\-\- line-cores and the continuum, 
similar to the $S$ index.

\section{Mg \scriptsize{II} \normalsize h+k and Ca
  \scriptsize{II} \normalsize H+K emission  index calibration}\label{calindex}

To obtain the Mount Wilson index $S$ indirectly from
the UV  spectra, we analysed the relation between the Ca \II\- index  and the Mg \scriptsize{II} \normalsize
line-core flux. To guarantee that both diagnostics re\-present the same phase
of activity of the star, we used 117 nearly simultaneous observations
(i.e. with a  time interval lower than 36 hours) of Ca \II\- and Mg
\II\- line-core surface fluxes.

These observations of 21 stars with $0.45\le$$B-V$$\le1.00$ (see Table
\ref{tablest}) are included in the dataset used by
\cite{1992A&A...258..432S} to intercalibrate Ca \scriptsize{II}
\normalsize and Mg \scriptsize{II} \normalsize surface fluxes. We
excluded from Schrijver's list HD 188512 due to its noisy IUE spectrum
and some observations of HD 3651 where the ratio between the Mg
\scriptsize{II} \normalsize k to h line fluxes is greater than 1.55
(i.e. where the k/h ratio deviates in more than 2$\sigma$ from the
mean).

\begin{table}[htb!]
\caption{\textbf{Stars used in the Mg \scriptsize{II} \normalsize -
Ca
    \scriptsize{II}  \normalsize calibration. }}\label{tablest}
\centering
\begin{tabular}{r  c  r  r r}
\hline
Stars &  Spectral &       &      & T$_{eff}$\\
 HD &   Class and &  m$_v$& $B-V$    & (K)\\
    &   Type      &       &        &    \\
\hline
\hline
  1835 & G3V & 6.39 & 0.66 & 5675\\
  9562 & G0V & 5.76 & 0.64 & 5750\\
 10700 & G8V & 3.50 & 0.72 & 5500\\
 17925 & K0V & 6.04 & 0.87 & 5170\\
 20630 & G5V & 4.83 & 0.68 & 5610\\
 22049 & K2V & 3.73 & 0.88 & 5140\\
 26965 & K1V & 4.43 & 0.82 & 5295\\
 35296 & F8V & 4.99 & 0.53 & 6185\\
 39587 & G0V & 4.41 & 0.59 & 5950\\
 45067 & G0V & 5.87 & 0.56 & 6060\\
101501 & G8V & 5.33& 0.72 & 5500\\
114378 & F5V & 5.22 & 0.45 & 6540\\
115383 & G0V & 5.22 & 0.59 & 5950\\
115404 & K3V & 6.52 & 0.94 & 4990\\
131156 & G8V & 4.72 & 0.72 & 5500\\
141004 & G0V & 4.43 & 0.60 & 5910\\
143761 & G2V & 5.41 & 0.60 & 5910\\
149661 & K0V & 5.75 & 0.82 & 5295\\
152391 & G8V & 6.64 & 0.76 & 5295\\
154417 & F8V & 6.01 & 0.58 & 5985\\
187013 & F5V & 4.99 & 0.47 & 6445\\
\hline
\hline
\end{tabular}

\end{table}

\cite{1984A&A...130..353R} found a relation between the Ca \II\-
line-core surface flux and the Mount Wilson index $S$ for main sequence stars with $0.30\le B-V\le 1.60$
\begin{equation}
\centering
F_{Ca\II}=F_H+F_K=S\,C_{cf}\,T_{eff}^4\,10^{-14},
\label{rutten_form}
\end{equation}
where the conversion factor $C_{cf}$ is given by

\begin{displaymath}
\log(C_{cf})=0.25(B-V)^3-1.33(B-V)^2+0.43(B-V)+0.24,
\nonumber
\end{displaymath}
 and $F_{H}$ and
$F_{K}$ are the Ca \II\- H and K surface fluxes expressed in
arbitrary units, which differ from an
absolute calibrated scale by a multiplicative factor
1.29$\times$10$^6$ erg cm$^{-2}$ s$^{-1}$. 

\begin{figure}[h!]
\resizebox{\hsize}{!}{\includegraphics{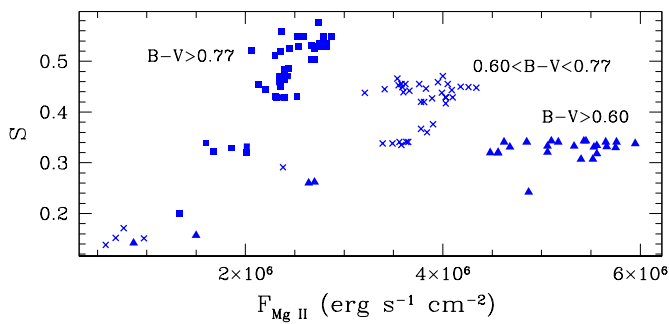}}
\caption{$S$ vs. $F_{MgII}$ for three different colour bins:
    \mbox{$B-V>0.77$} (squares), \mbox{$0.60<B-V<0.77$}
    (crosses) and \mbox{$B-V<0.60$} (triangles).}\label{calibbv}
\end{figure}

From the $F_{Ca\II}$ values listed in \cite{1992A&A...258..432S}, we obtained
the Mount Wilson $S$  using Eq. \ref{rutten_form}. On the other hand, we
measured the Mg \scriptsize{II} \normalsize line-core flux $F_{Mg\II}$ on the
corresponding 117 high resolution IUE spectra. The results are shown in Fig. \ref{calibbv}.

The main source of error in our data is the dispersion in $F_{Mg\II}$,
due to the fact that there is  a single value of $F_{Ca \II}$
for IUE spectra differing in less than 36 hours. We
found that the standard deviation of these $F_{Mg\II}$ values could
be up to  10\%. Therefore, we used in the calibration an average of the
$F_{Mg\II}$ values which were assigned a single $F_{Ca\II}$ and an
error of 10\% for $F_{Mg\II}$. Our dataset was
consequently reduced to 93 points.  

In Fig. \ref{calibbv} it can be seen that, as colour increases,
the slope of the relation between the Mg \scriptsize{II} \normalsize flux 
and the $S$ index also increases.   
This could be due to different reasons: on one hand, the 
Mount Wilson index calculation involves the
continuum flux near the Ca \II\- lines, which of course depends on $B-V$. On the other hand, a colour dependent 
basal flux, independent from the activity level of the star, could
also be present in the Mg \II\- and the Ca \II\- line-core emission.  
Many studies (e. g. \citealt{1989ApJ...341.1035S}, \citealt{1991A&A...252..203R})
showed that the basal flux in the chromospheric lines decreases as $B-V$ 
increases and that the  Mg \II\- basal flux is lower than the Ca \II\-
one. 

\begin{figure}[htb!]
\resizebox{\hsize}{!}{\includegraphics{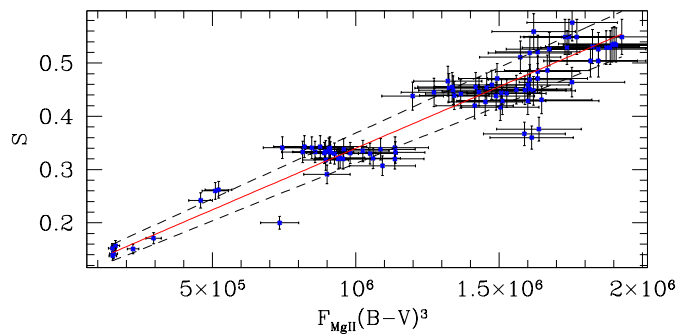}}
\caption{S vs. $F_{Mg II}$($B-V$)$^3$ for the stars listed in Table \ref{tablest}. The errors are assumed
  to be 10\% for $F_{Mg\II}$ and 6\% for $S$. The least
  square fit  (solid line) has a correlation coefficient of 0.95.  The
  dotted lines indicate $\pm$3$\sigma$ from
  the fit.}\label{S_fmg_alpha}
\end{figure}

Therefore, a colour dependent $S$\,-$F_{MgII}$ calibration is needed for these
data. Here, we proposed
an $S$-$F_{MgII}$ calibration given by
\vspace{-0.1cm}
\begin{equation}
S = a\,(B-V)^\alpha\,F_{Mg II}+b\,,\label{eq.calibfmg}
\end{equation}
whith  $F_{Mg\II}$ expressed in
erg cm$^{-2}$ s$^{-1}$, and we found that the best correlation coefficient (R=0.95) is obtained for
$\alpha$=(3.0 $\pm$ 0.1). Considering the errors in both coordinates  and
minimizing the expression
\begin{equation}
\centering
\chi^2=\Sigma_{i=1}^N\frac{(y_i-b-ax_i)^2}{\sigma_{y_i}^2+a^2\sigma_{x_i}^2},
\label{ec.chi_xy}
\end{equation}
where $y\equiv S$ and $x\equiv F_{Mg\II}(B-V)^\alpha$, we obtained
that the best parameters are \mbox{$a=(2.310\pm0.052)\times 10^{-7}$}
and \mbox{$b=0.109\pm 0.004$}, whith a reduced
$\chi^2=1.07$ for an uncertainty of 6\% in the $S$ values. 
In Fig. \ref{S_fmg_alpha} we plot the Mount Wilson index
$S$ and the Mg \scriptsize{II} \normalsize line-core surface flux
corrected  in colour and the best linear fit given by
Eq. \ref{eq.calibfmg}.

\section{Application of the calibration}\label{appl}

As an application of the  calibration obtained in
 Eq. \ref{eq.calibfmg}, we have measured the  Mg \scriptsize{II}
 \normalsize  line-core flux on all the 1623 IUE high resolution
spectra available for main sequence stars of spectral types F to K, and then
 converted the
 surface flux $F_{Mg II}$ to the Mount
 Wilson index. 
These results are presented in Table \ref{tabla_total}.

\begin{table*}[h!]
\begin{minipage}{\textwidth}
\caption{\textbf{ Long and short term variability records of the stars most observed
 by IUE.}}\label{timescalevar}
\begin{tabular}{r l c c c  c c l c l }
\hline
  &   & & & & &Long  scale\footnote{Maximum and minimum level of activity reached along
decades.} & & Short scale\footnote{Variations recorded in particular years.}  &   \\
   &  & && Approx.\footnote{References for age: 1. \cite{2002A&A...393..225M}, 2. \cite{1998A&A...330.1077D},
 3. \cite{1999A&A...348..897L}, 4. \cite{2005A&A...443..609S},
 5. \cite{2004A&A...418..989N} , 7. \cite{2000ApJ...531..503G} and
 8. \cite{1998A&A...338..455F}.} &Mean &variability  & & variability &   \\

Star &Sp. type& &$P_{rot}$\footnote{References for $P_{rot}$: a. \cite{1996ApJ...466..384D},
b. \cite{1997ApJ...483..947G} , c. \cite{1997MNRAS.284..803S},
d. \cite{1991ApJ...372..610H},  e. \cite{1997AAS...18912004J} and
f. \cite{1998ApJ...502..918B}.}& age & activity &($\sim$ decades)& Period of&($\sim$ month)&\\ 
HD &\& class  &$B-V$ & (days)& (Gyr)& $\langle S\rangle$\footnote{Mean
  annual Mount Wilson index.}&$S_{min}$/$S_{max}$& time& $S_{min}$/$S_{max}$&Year\\
\hline
\hline
 1835 &G3V &0.67 &7.78$\,^\textrm{a}$&0.6$\,^\textrm{\scriptsize{1}}$ &0.334 &0.315/0.353 &1981-2005 &0.307/0.355&1991\\
 2151 &G2IV &0.62& 28.00$\,^\textrm{b}$ &6.7$\,^\textrm{\scriptsize{2}}$ &0.153 & 0.133/0.163 & 1978-1995 &0.160/0.183&1985\\
        &  & &   &  &  &      &  &0.141/0.171&1993\\
        &  & &   &  &  &      &  &0.145/0.164&1994\\
10700& G8V &0.72&34.50$\,^\textrm{c}$&7.2$\,^\textrm{\scriptsize{3}}$  &0.175 & 0.171/0.186&1983-2004&0.167/0.173 & 1993\\
      &    &    &   & &             &  &       &0.172/0.178& 2002\\
       &    &    &   &     &        &   &      &0.172/0.181& 2003\\
      &    &    &   &      &       &    &     &0.169/0.179& 2004\\
 20630 & G5V& 0.68& 9.24$\,^\textrm{a}$ &0.7$\,^\textrm{\scriptsize{1}}$  &0.370 & 0.334/0.419 &1980-1994&0.357/0.382&1985\\
       &    &     &  & &   &     & & 0.318/0.364 &1994\\       
   22049 & K2V &0.88&11.68$\,^\textrm{a}$&0.7/0.9$\,^\textrm{\scriptsize{4}}$ &0.473 & 0.426/0.520  & 1978-2005&0.449/0.504&1980\\
         & &   &     &&               &     & &  0.395/0.460&1982\\    
         & &   &     &&               &     & &  0.495/0.519&1983\\
         & &   &     &&               &     & &  0.422/0.475&1986\\
         & &   &     &&               &     & &  0.423/0.459&2002\\
         & &   &     & &              &     & &  0.450/0.569&2003\\
    39587&G0V &0.59 &5.89$\,^\textrm{a}$&5.6$\,^\textrm{\scriptsize{5}}$ &0.335 &0.180/0.375 &1978-1993 &0.327/0.358  &1984\\
   115383 &G0V  &0.59 &3.33$\,^\textrm{a}$&5.1$\,^\textrm{\scriptsize{5}}$ &0.332 &0.248/0.362& 1978-1995& 0.296/0.375&1993\\
   128620 &G2V &0.71&29.00$\,^\textrm{d}$&6.8/7.6$\,^\textrm{\scriptsize{7}}$ &0.167 &0.153/0.185  & 1978-2004 &0.133/0.192   &1995\\
 128621 &K1V &0.88&36.90$\,^\textrm{e}$&6.8/7.6$\,^\textrm{\scriptsize{7}}$ &0.214 &0.162/0.248  &1978-2004 & 0.216/0.247&1978\\
         &  &  &    & & &     & &0.192/0.230 &1980\\
         &  &  &    & & &     & & 0.167/0.229&1982\\
         &  &  &    & & &     & & 0.206/0.225 &1983\\
         &  &  &    & & &     & & 0.169/0.297&1995\\
   131156 &G8V&0.76 &6.31$\,^\textrm{a}$& 2$\,^\textrm{\scriptsize{8}}$ &0.430 & 0.242/0.508 &1978-1993 &0.258/0.492 &1978\\
         &    & &    & & &     & &0.327/0.508 &1982\\
         &    & &    & & &     & &0.413/0.431 &1985\\
         &    & &    & & &     & &0.442/0.469 &1986\\
         &    & &    & & &     & & 0.407/0.473&1993\\
   133640 &G0Vnvar &0.65&0.28$\,^\textrm{f}$ &15.4$\,^\textrm{\scriptsize{5}}$ &0.272 &0.263/0.276  &1978-1990 &0.241/0.307 &1979\\
         &     & & &  &   &  & & 0.256/0.297&1989\\
\hline
\hline
\end{tabular}\\
\end{minipage}
\end{table*}

For several of the most observed main sequence stars, we analysed the
Mount Wilson index $S$ inferred from the IUE spectra,
together with the ones obtained from CTIO spectra
\citep{1996AJ....111..439H} and from CASLEO spectra  with the
calibration of \cite{2007astro.ph..3511C}. In this way our
observations cover the period between 1978 and 2005.  Even if, for
most stars, the density of measurements along the years is low, we can
infer the level of activity and its variability for the
whole period and during short intervals of time. For these stars, we
list in Table \ref{timescalevar} the average, maximum and
minimum level of activity reached along decades (columns 4 and 5) and the
variations recorded in particular years (column 6).

For those stars in Table \ref{timescalevar} that were observed for
decades and for which we have a large number of measurements, in what
follows we plot the
index $S$ vs. time and we analyse their magnetic behaviour in
detail.

\subsection*{HD 1835 - BE Ceti}

\begin{figure}[htb!]
\centering
\includegraphics[width=0.5\textwidth]{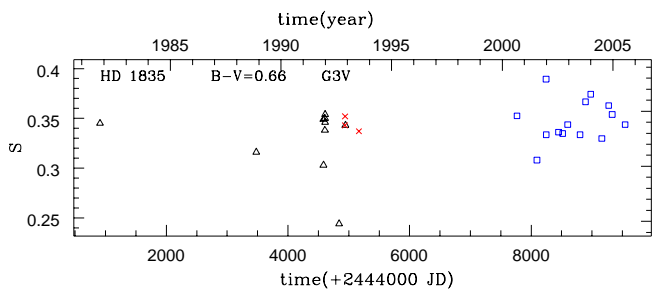}
\caption{Data for HD 1835. We indicate the Mount Wilson index $S$
obtained from the IUE spectra with triangles ($\triangle$), the one
obtained by \cite{1996AJ....111..439H} from CTIO spectra with crosses
($\times$) and the one obtained from CASLEO spectra with square dots
($\Box$).}\label{hd1835}
\end{figure}

 As expected, the BY Dra stars (HD 1835 and HD 22049) show very strong
 chromospheric activity. In particular, HD 1835(G3V) is a
 young star only
 600 Myr old, with  a rotation period of 7.78 days (see references in
 Table \ref{timescalevar}). 
\cite{1995ApJ...438..269B}, studying data for
 the period 1966-1991, reported that it  presents an
 activity cycle with a length of $9.1\pm 0.3$ years. In
 Fig. \ref{hd1835} we show our data for this star. Unfortunately, we
 are not able to check this period from these data, as they are
 insufficient to build a periodogram and obtain a significant
 result. However, we can infer that the level of activity slightly increased in
 the period 2000-2005 with respect to the previous years.  From
 Fig. \ref{hd1835} we obtain that HD 1835 reached a maximum level of
 activity $S_{max}$=0.354, a minimum level $S_{min}$=0.244 and a mean
 level $\langle S \rangle=$0.332 between 1978 and 1995, while between
 2000 and 2004 these values were $S_{max}$=0.389, $S_{min}$=0.308 and
 $\langle S \rangle=$0.347.

\subsection*{HD 22049 - $\epsilon$ Eri}

HD 22049 ($\epsilon$ Eri, K2V)  is a young  and
  chromosperically active star $\sim$0.8 Gyr old with a rotation period of 11.28 days (see references in
  Table \ref{timescalevar}). 
\cite{1995ApJ...438..269B} reported that HD 22049  is a
variable star without an evident cyclic behaviour, but
\cite{1995ApJ...441..436G} reported an underlying magnetic cycle with a period of the order of 5
years from 1986 to 1992. 
  \cite{2007AJ....133..862H}  found that $\epsilon$ Eri is a high-activity variable star with a mean
  level of activity given by $\langle S \rangle=0.516$ between 
  1994 and 2005. From our data, we obtained a  $\langle S \rangle=0.479$ for this time interval, which
  coincides with \citeauthor{2007AJ....133..862H}'s results within the
  statistical error.

\begin{figure}[htb!]
\centering
\includegraphics[width=0.5\textwidth]{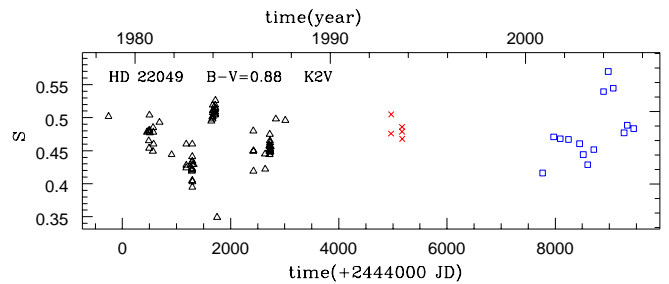}
\caption{Data for HD 22049. Symbols as in Fig. \ref{hd1835}.}\label{hd22049}
\end{figure}

 In
Fig. \ref{hd22049} we show our data for this star, where we observe appreciable short-scale ($\sim$ months)
variations from 10\% to 25\%.

\begin{figure}[htb!]
\centering
\subfigure[\label{hd22049_per}]{\includegraphics[width=0.23\textwidth]{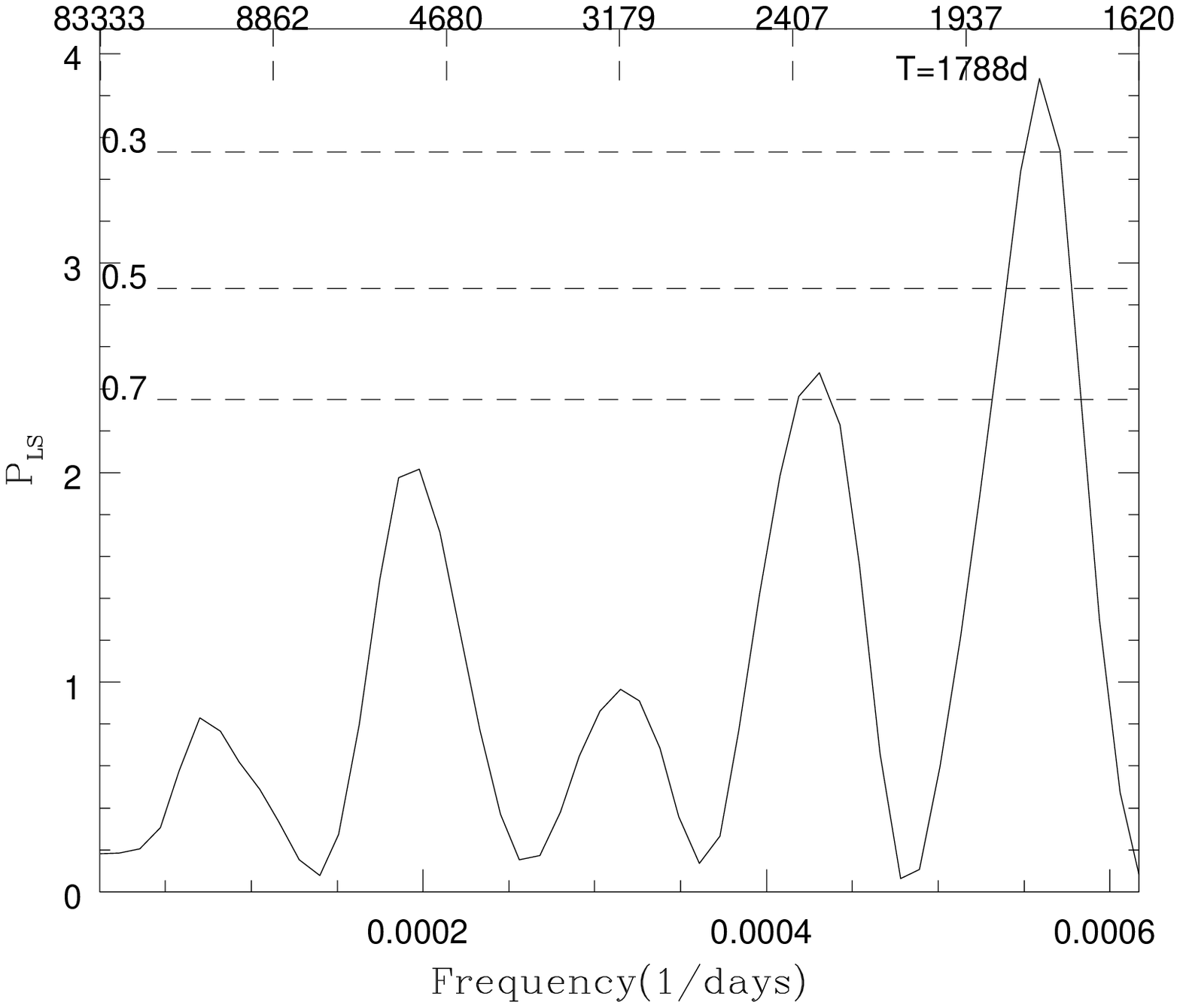}}\hfill
\subfigure[\label{hd22049_fas_prom}]{\includegraphics[width=0.23\textwidth]{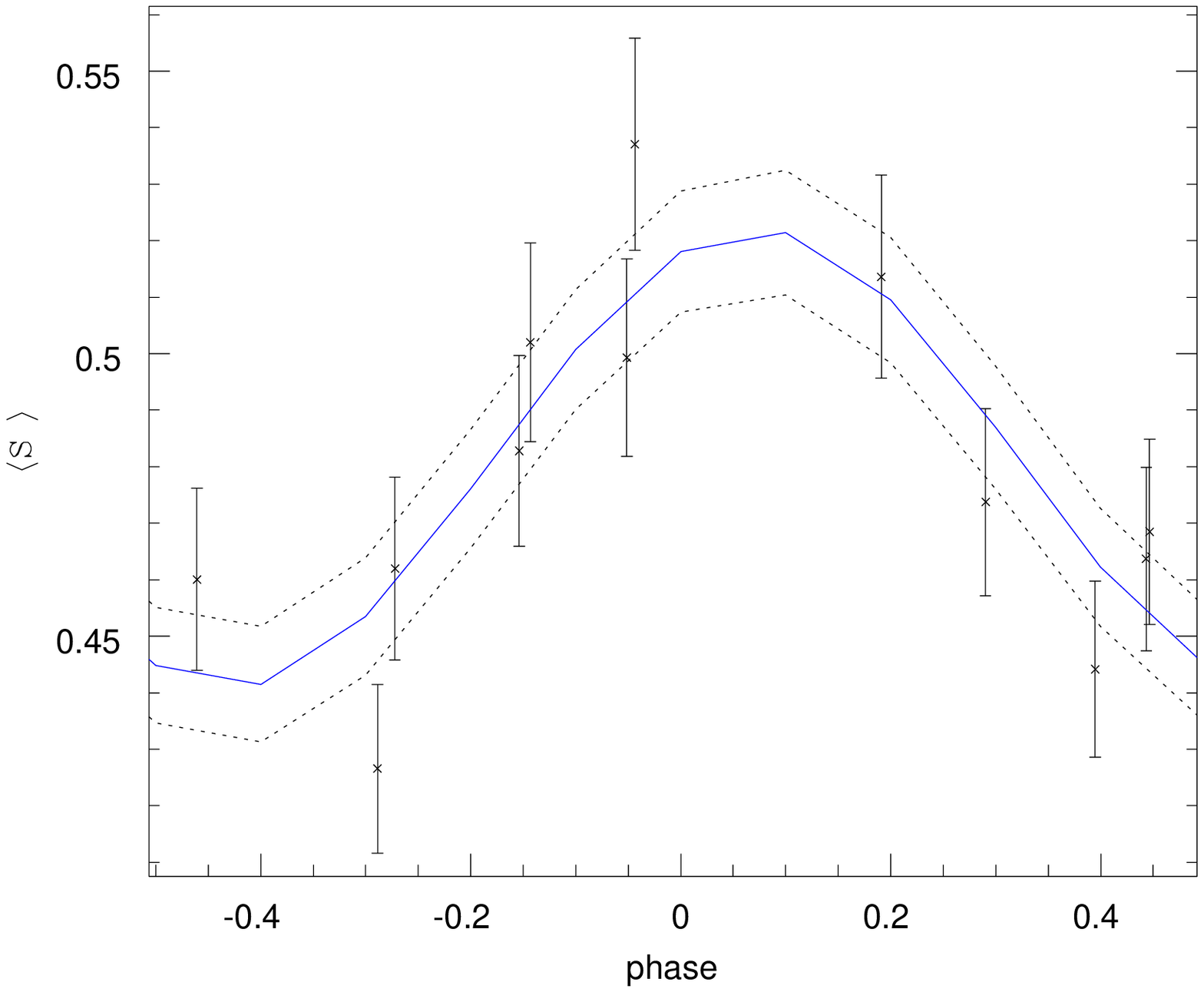}}
\caption{HD 22049.  \emph{Left}: Lomb-Scargle periodogram of the mean annual
$\langle S \rangle$ of the data plotted in
Fig. \ref{hd22049}. The False Alarm Probability (FAP)
levels of 30, 50 and 70\% are indicated. \emph{Right}: The mean annual $\langle S \rangle$
  of the data plotted in Fig. \ref{hd22049} phased with the period
  of 1788 days. The solid line shows the harmonic curve that best fits the
  data with a 60\% confidence level and the dashed lines indicate the points
  that apart $\pm 3\sigma$ from that fit. }
\end{figure}

 For our time interval (1978-2004), we
analysed the mean annual $\langle S\rangle$ as a function of time with
the Lomb-Scargle periodogram (\citealt{1982ApJ...263..835S},
\citealt{1986ApJ...302..757H}), using the algorithm given by
\cite{1992nrfa.book.....P}. This periodogram is shown in
Fig. \ref{hd22049_per}, where it can be seen that there is, indeed, a
peak at 1788 days ($\sim$4.9 years), with a false alarm probability
(FAP) of 22\%. In Fig. \ref{hd22049_fas_prom}, we plot the mean anual
$\langle S \rangle$ phased with
the period obtained (1788 days) and we found that an harmonic function
fits these points with a 60\% confidence level.

\subsection*{HD 10700 - $\tau$ Ceti}

 HD 10700 ($\tau$ Ceti, G8V) is an old slow rotator (see Table \ref{timescalevar}), known to be rather constant in activity
\citep{1994ApJ...427.1042G}. In particular, \cite{1995ApJ...438..269B} analysed the
Mount Wilson index from 1966 to 1991, and found that the chromospheric
activity of this star is almost flat with a mean $\langle S\rangle=0.171$.
They proposed that it could be in a magnetic
phase analogous to the solar Maunder minimum. They also found that HD
10700 is possibly increasing its level of activity since 1989.

\cite{2004ApJ...609..392J} compared solar UV spectra with several STIS
spectra of HD 10700 from \emph{Hubble Space Telescope} obtained in
August, 2000, and concluded that its line spectrum may represent a solar
spectrum corresponding to the Maunder minimum. On the other hand, they
also found evidence of a weak magnetic field in HD 10700.  Therefore,
\cite{2004ApJ...609..392J} proposed that this star has temporally
slipped into an extended magnetic quiescence (like the solar Maunder
minimum) and has only occasional active regions because of a
small-scale turbulent dynamo.
\begin{figure}[htb!]
\centering
\includegraphics[width=0.5\textwidth]{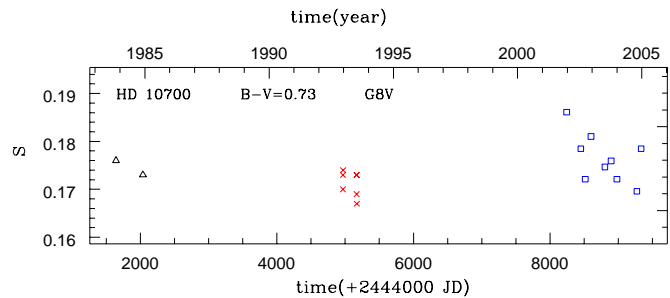}
\caption{ Data for HD 10700. Symbols
as in Fig. \ref{hd1835}. }\label{hd10700}
\end{figure}

  From the
few data plotted in Fig. \ref{hd10700}, we observe  that the mean level
of activity of HD 10700 seems to increase in 2001. In particular  we
obtained a
mean $\langle S \rangle=0.172\pm 0.001$ between 1978 and 1995, in agreement with the value given by \cite{1995ApJ...438..269B}, and $\langle S
\rangle=0.176\pm0.003$ between 2001 and 2005. We also note a
$\sim$10\% variation from maximum to minimum between the end of 2001
and 2002, which  could be attributed to
faint magnetic events.
 
In summary, we find that HD 10700 could have increased its level
 of activity since 2001, and that its chromospheric variability in the
 period 2001-2005 is also appreciable. Therefore, our results are in
 agreement with the model proposed by \cite{2004ApJ...609..392J} of an
 underlying turbulent dynamo responsible for the few signs of magnetic
 activity in this star.

\subsection*{HD 2151 - $\beta$ Hyi }

\begin{figure}[htb!]
\centering
\includegraphics[width=0.5\textwidth]{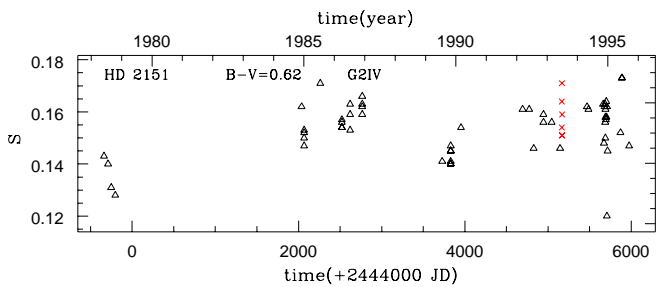}
\caption{Data for HD 2151 ($\beta$ Hyi). Symbols as in Fig. \ref{hd1835}.}\label{hd2151}
\end{figure}

 \cite{1993ApJ...403..396D} affirm that HD 2151 ($\beta$ Hyi, G2IV)
can be considered an evolved Sun, with a well-determined age of
$\sim$6.7 Gyr \citep{1998A&A...330.1077D}, and a magnetic activity
cycle probably longer than the solar one (15-18 years). In
Fig. \ref{hd2151} we plot our data for this star. With these data, we
obtain  a mean
chromospheric variation of 22\% from 1978 to 1995. On the other hand,
the short-scale variations registered in 1985, 1993 and 1994 do not
exceed 17\%.

\begin{figure}[htb!]
\centering
\subfigure[\label{hd2151_per}]{\includegraphics[width=0.23\textwidth]{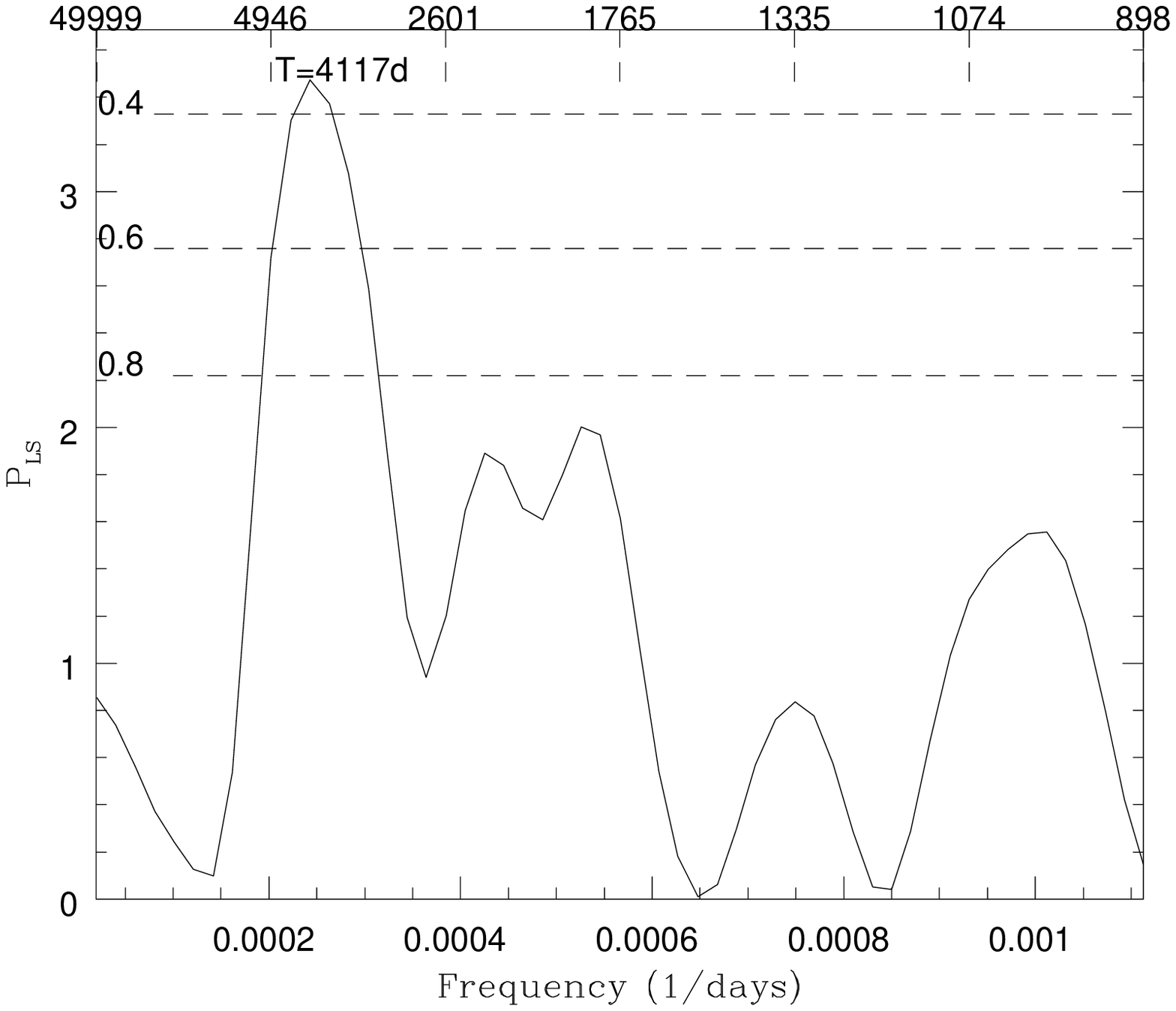}}\hfill
\subfigure[\label{hd2151_fas_prom}]{\includegraphics[width=0.23\textwidth]{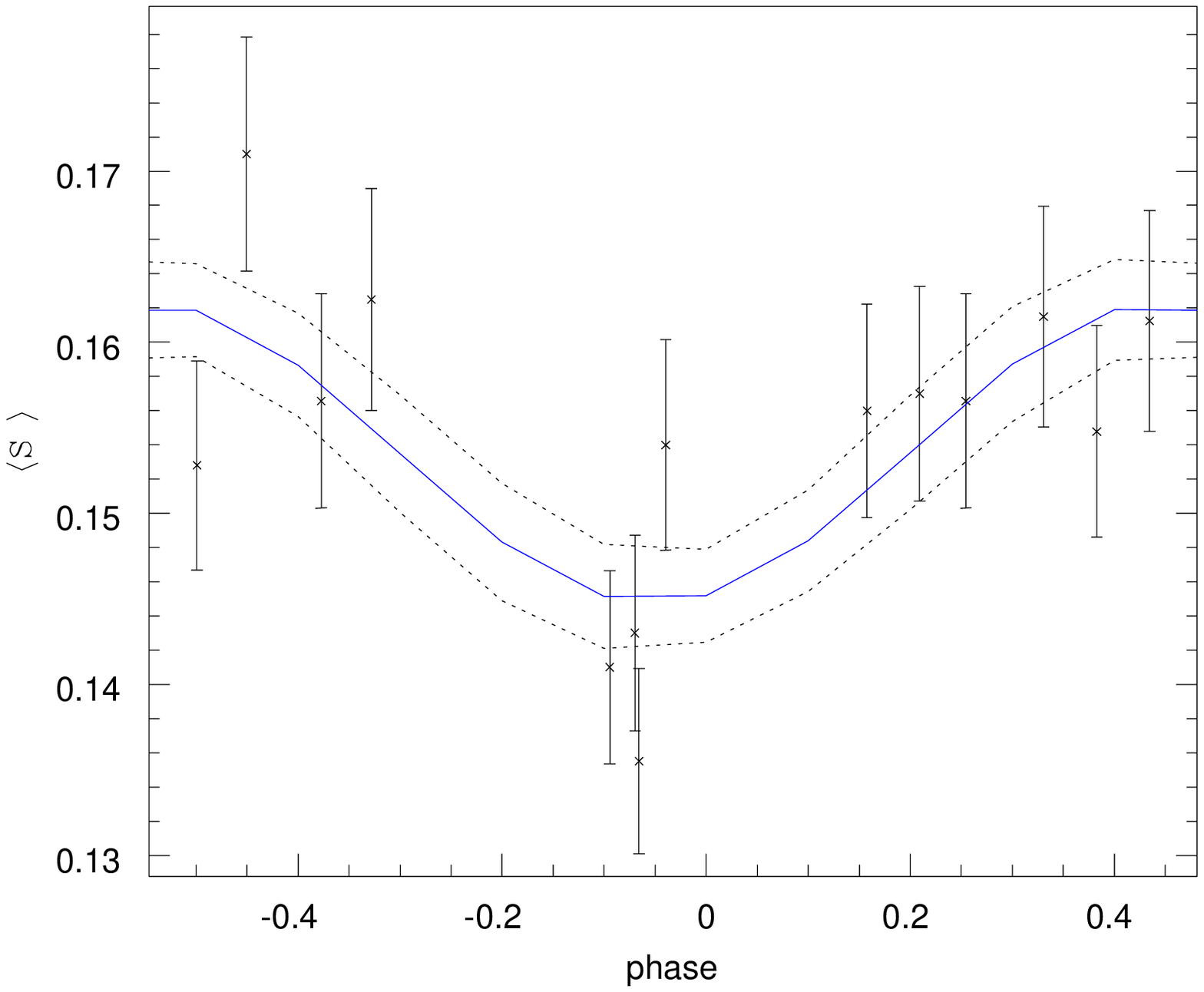}}
\caption{HD 2151 ($\beta$ Hyi). \emph{Left}: Lomb-Scargle periodogram  of
 the mean annual $\langle S \rangle$ in Fig. \ref{hd2151}. The False Alarm Probability levels of 40 to 80\%
are  indicated. \emph{Right}: The mean annual $\langle S \rangle$
  of the data plotted in Fig. \ref{hd2151} phased with the period of
   4117 days.  The solid line represents the harmonic curve that best fits the
  data with a 50\% confidence level and the dashed lines indicate the points
  that appart $\pm 3\sigma$ from that fit.}
\end{figure}

 To explore the magnetic
behaviour of this star, we analysed the mean annual $\langle S\rangle$ with the Lomb-Scargle algorithm, and we
obtained a cyclic behaviour with a  period of 4117 days ($\sim$11.28
years), with  a FAP of 35\%. The periodogram is plotted in
Fig. \ref{hd2151_per}. In
Fig. \ref{hd2151_fas_prom}  we show the mean annual $\langle S \rangle$ phased with this
period and the harmonic curve that best fits the data with a good
confidence level of 50\%. Recently,  \cite{2007MNRAS.tmpL..50M} also  obtained a $\sim$12 year-magnetic
activity cycle for HD 2151 by analysing a Mg \II\- index derived from
the same IUE data plotted on Fig. \ref{hd2151} and they also found a
faint correlation between this activity cycle and astereosismic
observations.

\subsection*{HD 128620 - $\alpha$ Cen A}  
The non-interactive 79.2 yr  binary system $\alpha$ Centauri is composed by the
G2V star HD 128620 ($\alpha$ Cen A) and the K1V star HD 128621
($\alpha$ Cen B), with a relative distance of 23.50 AU
\citep{1978AJ.....83.1653K}.  Both stars have different levels of
activity. From 1978 to 2004, HD 128620 presented a mean annual index
$\langle S\rangle$=0.167 while HD 128621 presented  $\langle
S\rangle$=0.214. 

\begin{figure}[htb!]
\centering
\includegraphics[width=0.5\textwidth]{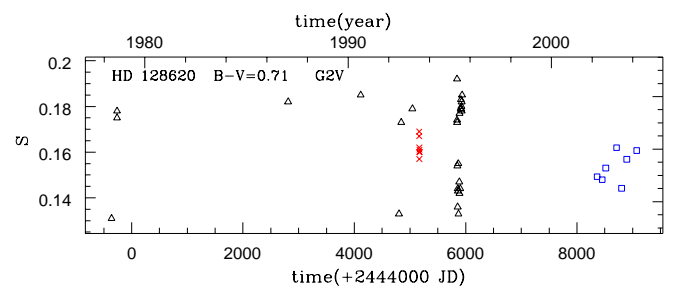}
\caption{Data for HD 128620. Symbols as in Fig. \ref{hd1835}.}\label{hd128620}
\end{figure}

$\alpha$ Cen A is considered a good solar-twin (G2V, $\sim$ 1.09 M$_\odot$,
solar abundances).  In particular, \cite{1992A&A...256..185C} reported
that it has a UV spectrum similar to the inactive Sun, and
\cite{2004ApJ...609..392J} concluded that HD 128620 provides a good
proxy for the Sun's UV spectrum during an intermediate phase of the
solar activity cycle.  From our data, which is plotted in
Fig. \ref{hd128620}, we found that the mean annual $\langle S \rangle$
presented an 11\% variation along 16 years (1978-2004), which is in
agreement with the analogy of HD 128620 with a Sun of moderate
activity. However, in Table \ref{timescalevar}
we observe an appreciable short-scale chromospheric
variation ($\sim$44\%) in the index $S$ during 1995, much larger than
what is observed in the Sun. Therefore, HD 128620 probably presents larger
chromospheric features than the Sun.

Recently, \cite{2005A&A...442..315R} found that HD 128620 has
  decreased its X-ray flux by at least an order of magnitude during their
observation program (from March 2003 to February 2005), and they
attributed this variation to an irregular event or an unknown coronal
activity cycle.  To search for a magnetic cycle analogous to the solar one, we
analysed the mean annual $\langle S \rangle$ of the data plotted in
Fig. \ref{hd128620} with the Lomb-Scargle periodogram, but we did not
find a periodic behaviour.  

\subsection*{ HD 128621 - $\alpha$ Cen B}

\begin{figure}[htb!]
\centering
\includegraphics[width=0.5\textwidth]{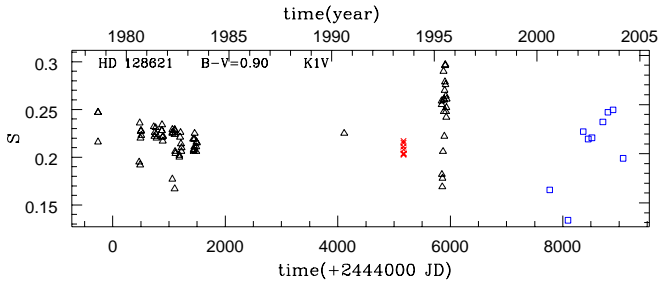}
\caption{Data for HD 128621. Symbols as in Fig. \ref{hd1835}.}\label{hd128621}
\end{figure}

\begin{figure}[htb!]
\centering
\includegraphics[width=0.45\textwidth]{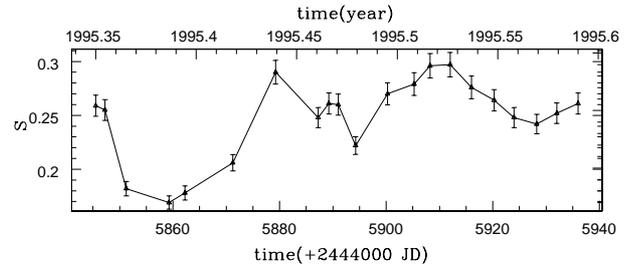}
\caption{Detail of Fig. \ref{hd128621} in 1995, showing a 35
  day-modulation in the light-curve.}\label{fig.hd128621_det}
\end{figure}

From ROSAT observations, \cite{2004A&A...417..651S} reported HD 128621
as a flare star as it presented variations of nearly 30\% in its X-ray
emission during 20 days in August 1996. The data for this star are
shown in Fig. \ref{hd128621}, where it can be seen that, during 1995,
it presented a chromospheric variation close to 75\%, which could be
attributed to a flare-like process.  However, the light-curve for this
event, which is shown in Fig. \ref{fig.hd128621_det}, does not present
the typical characteristics of flares.

To study if this
variation is due to rotational modulation, we analysed the data in
Fig. \ref{fig.hd128621_det} with the Lomb-Scargle periodogram and we
obtained a period of 35.1 days, similar to the values that can be found
in the literature for the rotation period. In particular,  \cite{1997MNRAS.284..803S}
estimated a rotation period of 42 days for HD 128621, while
\cite{1997AAS...18912004J} obtained a value of $\sim$37
days. Therefore, the variation found in Fig. \ref{fig.hd128621_det} is
probably due to rotation, for which we obtain a period of 35.1 days.

\begin{figure}[htb!]
\centering
\subfigure[\label{hd128621_per}]{\includegraphics[width=0.23\textwidth]{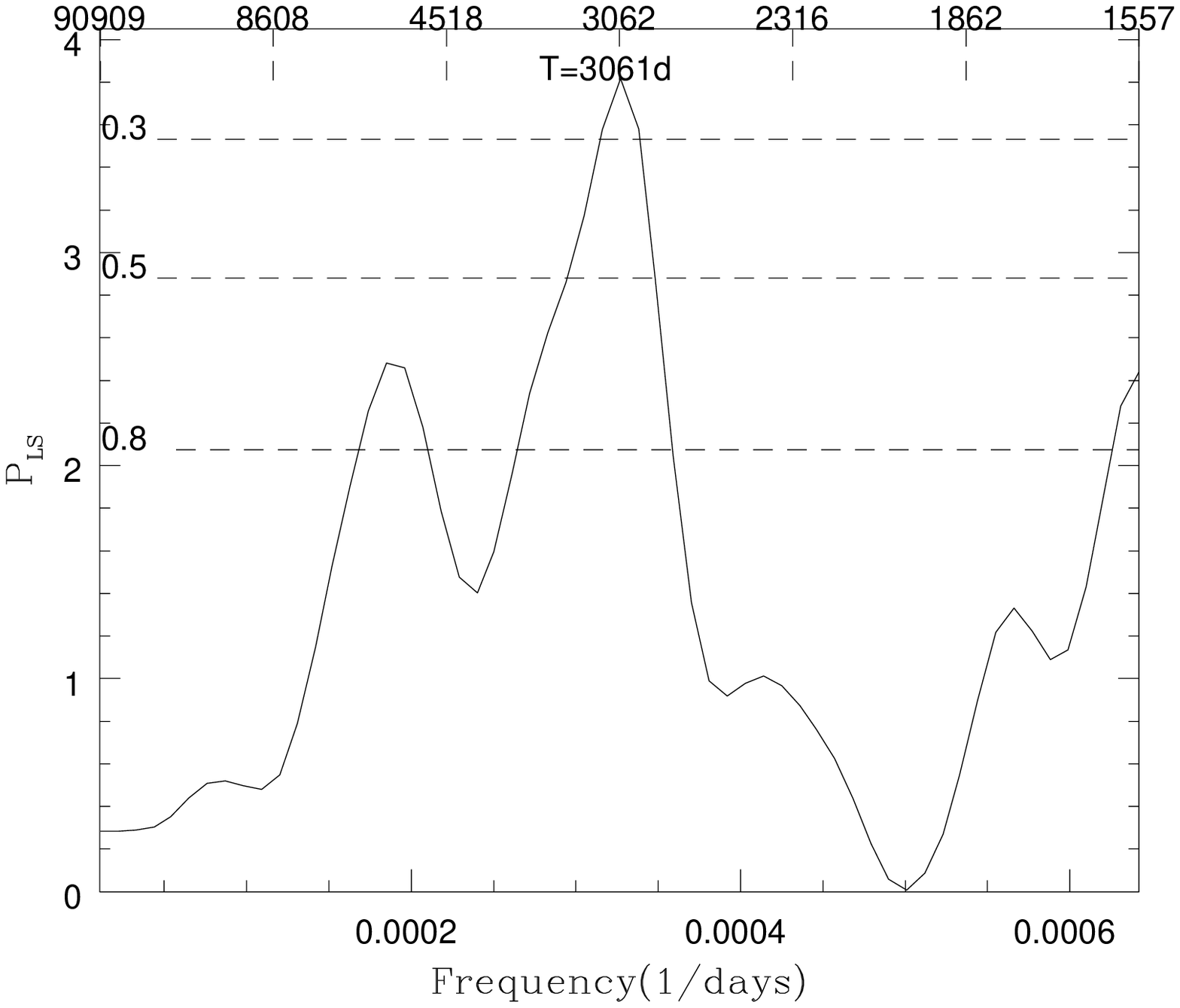}}\hfill
\subfigure[\label{hd128621_fas_prom}]{\includegraphics[width=0.23\textwidth]{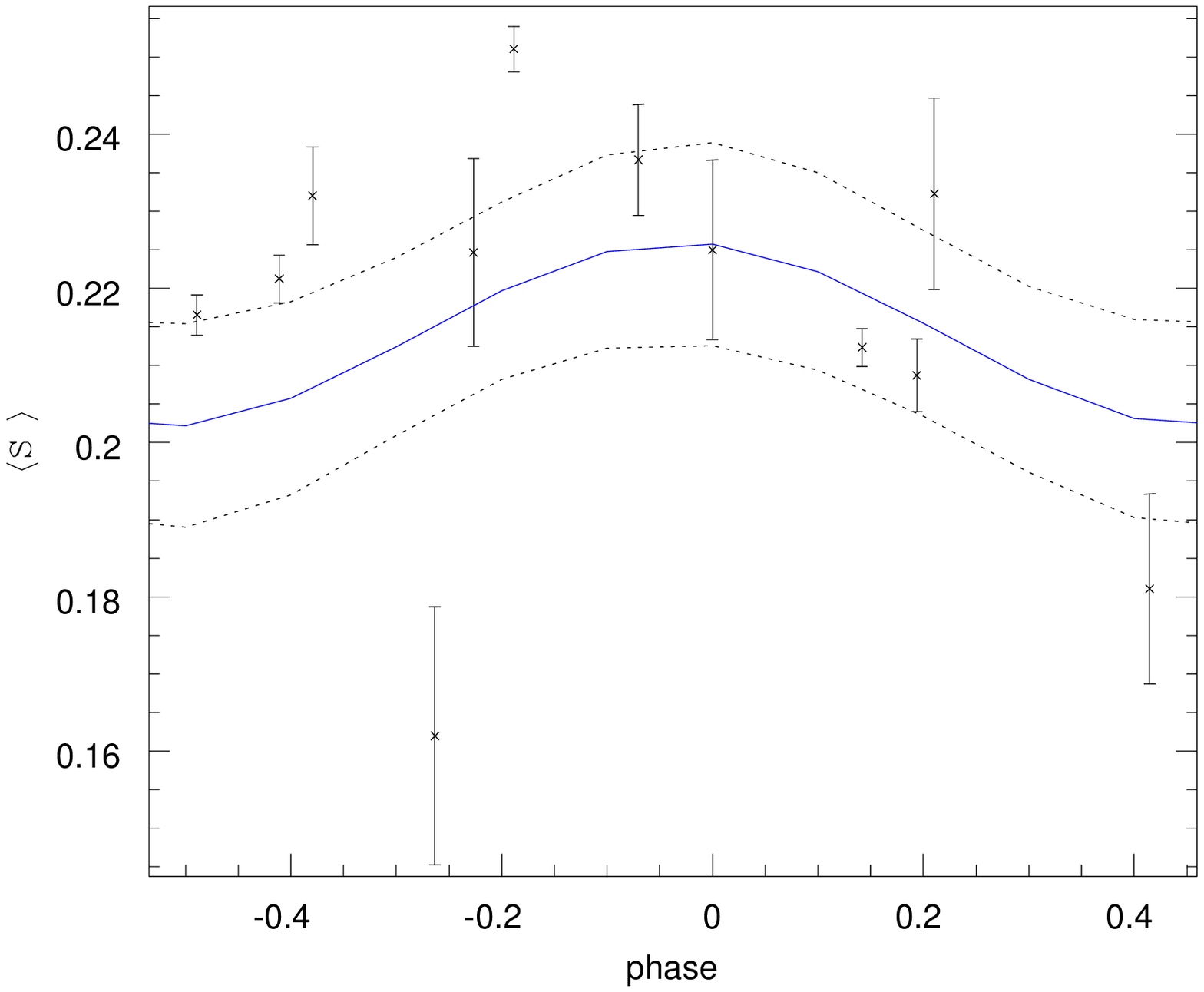}}
\caption{HD 128621.\emph{Left}: Lomb-Scargle periodogram of the mean
 annual $\langle S \rangle$ in Fig. \ref{hd128621}. The False Alarm
 Probability levels of 30, 50 and 80\% are 
 indicated. \emph{Right}: The mean annual $\langle S \rangle$ of the
 data plotted in Fig. \ref{hd128621} phased with the period of 3061
 days. The solid line represents the harmonic curve that best fits the
 data with an 80\% confident level and the dashed lines indicate the points
 that apart $\pm 3 \sigma$ from that fit.}
\end{figure}
To analyse the magnetic behaviour of HD 128621, we studied the mean
annual $\langle S\rangle$ of the data plotted in Fig. \ref{hd128621}
with the Lomb-Scargle periodogram, shown in
Fig. \ref{hd128621_per}. We obtained a magnetic cycle with a period of
3061 days ($\sim$ 8.38 years) with a FAP of 24\%. In
Fig. \ref{hd128621_fas_prom}, we also show the $\langle S \rangle$
phased with this period and the harmonic curve that best fits these
points with a confidence level of 80\%. The point for $\langle
  S \rangle\sim 0.16$, which significantly deviates from the harmonic curve, corresponds to the only registry of activity we have for the year
2000 and it is, therefore, not statistically representative of the
mean annual activity.

Our results are consistent with the decline in X-ray  lumonisity after 2005 reported by \cite{2007MmSAI..78..311R}.

\subsection*{HD 131156 A -  $\xi$ Boo}
\begin{figure}[htb!]
\centering
\includegraphics[width=0.5\textwidth]{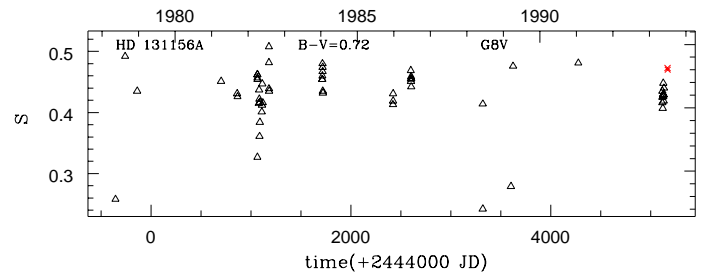}
\caption{ Data for HD 131156A. Symbols as in Fig. \ref{hd1835}.}\label{hd131156}
\end{figure}

HD 131156A is another flare star, which belongs to the visual binary
 system $\xi$ Boo. It presented chromospheric variations of 55\% during 1982
 and of 90\% during 1978.  \cite{1995ApJ...438..269B} reported that HD 1311156A
 is a variable star without any evident cyclic behaviour. We analysed
 the mean annual $\langle S \rangle$ of the data plotted in
 Fig. \ref{hd131156} with the Lomb-Scargle periodogram, but we did not
 obtain a significant period. \cite{1995ApJ...438..269B} reported an $\langle
 S\rangle$=0.461 between 1966 to 1993 and we found an
 $\langle S\rangle$  7\% lower from 1978 to 1994 (Table
 \ref{timescalevar}). These values are similar within the standard
 deviation, a fact which supports our calibration.

\section{Summary and Conclusions}
The main purpose of this work is to incorporate the UV spectroscopic observations
available in the IUE archives, and, in particular, using the Mg \II\- h and k
lines, to the systematic studies of magnetic activity in solar-type
stars. This allows us to extend the temporal span covered with these
studies.

First, we analysed the ultraviolet continuum flux near the Mg \II\- lines, and  we
obtained a relation between the mean UV continuum flux and the colour
$B-V$ of the star. We also found that there is an activity
component in this continuum flux. Therefore, an activity index
constructed as the
ratio of the fluxes in the Mg \II\- line-cores and the continuum,
similar to the Mount Wilson $S$-index, is not the best tool for the analysis
of chromospheric activity, since part of the activity-related signal
cancels out.

Subsequently, we analysed the relation between the Mount Wilson index
and the Mg \II\- line-core fluxes for a set of 117 nearly simultaneous
observations of Mg \II\- and Ca \II\- fluxes of 21 F5 to K3 main sequence
stars.  We obtained that the relation between the Mount Wilson $S$-index
and the Mg \II\- fluxes depends on the stellar colour
(i.e. spectral type), and we found
the relation between the index $S$ and the Mg \II\- line-core flux and
the stars's $B-V$.

From this calibration, we computed the Mount Wilson $S-$index for all high
resolution IUE spectra of F, G and K main sequence stars, totaling 1623
spectra of 259 stars.

To study the evolution of activity levels for the most observed stars,
 we used these indices together with the Mount Wilson indices derived
 from several spectra in the visible wavelength range, obtained at CTIO \citep{1996AJ....111..439H} and at CASLEO
 \citep{2007astro.ph..3511C}. In this way the data cover the period
 between 1978 and 2005. 

For the most frequently observed stars of this sample, we
analysed the level of activity over decades of time
and also studied the short-scale variations.  In particular,
we analysed the data of each star with the Lomb-Scargle periodogram
searching for periodic patterns analogous to the solar cycle. We
confirmed that HD 22049 ($\epsilon$ Eri, K2V) and HD 2151
($\beta$ Hydri, G2IV) present chromospheric activity cycles
of $\sim$5 and $\sim$12 years, respectively. We also found evidence of
an activity cycle for HD 128621 ($\alpha$ Cen B, K1V), with a period
of $\sim$8 years, which, to our knowledge, has not been reported in
the literature.  In particular, from our registry of activity,
we obtained a rotation period of $\sim$35 days for HD 128621, similar
to the one reported by \cite{1997AAS...18912004J}. On the
other hand, we did not find a cyclic pattern in the periodogram of its
companion HD 128620 ($\alpha$ Cen A, G2V).

\begin{table*}
\begin{minipage}[t]{\textwidth}
\caption{\textbf{ Index $S$ derived  from
all the  high resolution spectra of F, G and K dwarf stars
observed by IUE. }}\label{tabla_total}
\begin{tabular}{rrrrrrrrr}
\hline\hline
\footnote{The full table is
available in electronic form at the CDS via \textsf{http://cdsweb.u-strasbg.fr/cgi-bin/qcat?J/A+A/} .} HD &  $m_V$  &
 $B-V$ & Plx(mas) &     Julian
 date\footnote{IUE spectrum date.} &   Year$^{\scriptsize{a}}$ &    $S$\footnote{Mount Wilson Index derived from IUE
 spectrum.} & $\sigma_S$\footnote{Mount Wilson Index standard deviation} &   IUE spectrum\\
\hline
   166 &  6.07 &  0.752 &   72.98 &  2445182.000 &  1982.58 &   0.404 &   0.0161 & LWR13815\\
   166 &  6.07 &  0.752 &   72.98 &  2445290.000 &  1982.87 &   0.467 &   0.0189 & LWR14642\\
   166 &  6.07 &  0.752 &   72.98 &  2445291.000 &  1982.87 &   0.447 &   0.0180 & LWR14654\\
   166 &  6.07 &  0.752 &   72.98 &  2445292.500 &  1982.87 &   0.451 &   0.0182 & LWR14663\\
   400 &  6.21 &  0.504 &   30.26 &  2445998.500 &  1984.81 &   0.156 &   0.0058 & LWP04632\\
   432 &  2.28 &  0.380 &   59.89 &  2444521.250 &  1980.76 &   0.159 &   0.0059 & LWR08974\\
   432 &  2.28 &  0.380 &   59.89 &  2447112.000 &  1987.85 &   0.157 &   0.0058 & LWP12076\\
   432 &  2.28 &  0.380 &   59.89 &  2444509.000 &  1980.73 &   0.159 &   0.0058 & LWR08895\\
   432 &  2.28 &  0.380 &   59.89 &  2447111.750 &  1987.85 &   0.161 &   0.0059 & LWP12074\\
   432 &  2.28 &  0.380 &   59.89 &  2447111.750 &  1987.85 &   0.159 &   0.0059 & LWP12068\\
   432 &  2.28 &  0.380 &   59.89 &  2447111.500 &  1987.85 &   0.161 &   0.0059 & LWP12067\\
   432 &  2.28 &  0.380 &   59.89 &  2447027.000 &  1987.63 &   0.160 &   0.0059 & LWP11431\\
   432 &  2.28 &  0.380 &   59.89 &  2447111.750 &  1987.85 &   0.165 &   0.0061 & LWP12070\\
   432 &  2.28 &  0.380 &   59.89 &  2444085.250 &  1979.58 &   0.138 &   0.0052 & LWR05214\\
   432 &  2.28 &  0.380 &   59.89 &  2446676.750 &  1986.67 &   0.163 &   0.0060 & LWP08997\\
   432 &  2.28 &  0.380 &   59.89 &  2446676.750 &  1986.67 &   0.161 &   0.0059 & LWP08996\\
   432 &  2.28 &  0.380 &   59.89 &  2447111.750 &  1987.85 &   0.162 &   0.0059 & LWP12073\\
   432 &  2.28 &  0.380 &   59.89 &  2447112.000 &  1987.85 &   0.164 &   0.0060 & LWP12075\\
   432 &  2.28 &  0.380 &   59.89 &  2447111.750 &  1987.85 &   0.165 &   0.0061 & LWP12071\\
   432 &  2.28 &  0.380 &   59.89 &  2447111.750 &  1987.85 &   0.164 &   0.0060 & LWP12069\\
   432 &  2.28 &  0.380 &   59.89 &  2447111.750 &  1987.85 &   0.163 &   0.0060 & LWP12072\\
   432 &  2.28 &  0.380 &   59.89 &  2446677.000 &  1986.67 &   0.164 &   0.0060 & LWP08999\\
   432 &  2.28 &  0.380 &   59.89 &  2446677.000 &  1986.67 &   0.163 &   0.0060 & LWP09000\\
   432 &  2.28 &  0.380 &   59.89 &  2446677.000 &  1986.67 &   0.167 &   0.0061 & LWP08998\\
   432 &  2.28 &  0.380 &   59.89 &  2446677.000 &  1986.67 &   0.168 &   0.0062 & LWP09001\\
   483 &  7.07 &  0.644 &   19.28 &  2447899.000 &  1990.02 &   0.276 &   0.0105 & LWP17097\\
   693 &  4.89 &  0.487 &   52.94 &  2445641.750 &  1983.83 &   0.136 &   0.0051 & LWP02201\\
   905 &  5.71 &  0.331 &   28.57 &  2444523.500 &  1980.77 &   0.173 &   0.0063 & LWR08992\\
  1581 &  4.23 &  0.576 &  116.38 &  2444748.250 &  1981.39 &   0.139 &   0.0052 & LWR10687\\
  1581 &  4.23 &  0.576 &  116.38 &  2444729.500 &  1981.34 &   0.138 &   0.0052 & LWR10520\\
  1581 &  4.23 &  0.576 &  116.38 &  2444770.500 &  1981.45 &   0.139 &   0.0052 & LWR10858\\
  1581 &  4.23 &  0.576 &  116.38 &  2445059.750 &  1982.25 &   0.146 &   0.0054 & LWR12915\\
  1581 &  4.23 &  0.576 &  116.38 &  2443710.500 &  1978.55 &   0.127 &   0.0049 & LWR01862\\
  1581 &  4.23 &  0.576 &  116.38 &  2443798.000 &  1978.78 &   0.126 &   0.0049 & LWR02621\\
  1581 &  4.23 &  0.576 &  116.38 &  2444762.500 &  1981.43 &   0.146 &   0.0054 & LWR10799\\
  1581 &  4.23 &  0.576 &  116.38 &  2443912.000 &  1979.10 &   0.152 &   0.0056 & LWR03699\\
  1581 &  4.23 &  0.576 &  116.38 &  2446038.000 &  1984.91 &   0.153 &   0.0057 & LWP04914\\
  1581 &  4.23 &  0.576 &  116.38 &  2443748.500 &  1978.65 &   0.136 &   0.0051 & LWR02191\\
\hline
\end{tabular}
\end{minipage}
\end{table*}


\begin{thebibliography}{45}
\expandafter\ifx\csname natexlab\endcsname\relax\def\natexlab#1{#1}\fi

\bibitem[{{Baliunas} {et~al.}(1995){Baliunas}, {Donahue}, {Soon}, {Horne},
  {Frazer}, {Woodard-Eklund}, {Bradford}, {Rao}, {Wilson}, {Zhang}, {Bennett},
  {Briggs}, {Carroll}, {Duncan}, {Figueroa}, {Lanning}, {Misch}, {Mueller},
  {Noyes}, {Poppe}, {Porter}, {Robinson}, {Russell}, {Shelton}, {Soyumer},
  {Vaughan}, \& {Whitney}}]{1995ApJ...438..269B}
{Baliunas}, S.~L., {Donahue}, R.~A., {Soon}, W.~H., {et~al.} 1995, \apj, 438,
  269

\bibitem[{{Boehm-Vitense}(1981)}]{1981ARA&A..19..295B}
{Boehm-Vitense}, E. 1981, \araa, 19, 295

\bibitem[{{Brickhouse} \& {Dupree}(1998)}]{1998ApJ...502..918B}
{Brickhouse}, N.~S. \& {Dupree}, A.~K. 1998, \apj, 502, 918

\bibitem[{{Cassatella} {et~al.}(2000){Cassatella}, {Altamore},
  {Gonz{\'a}lez-Riestra}, {Ponz}, {Barbero}, {Talavera}, \&
  {Wamsteker}}]{2000A&AS..141..331C}
{Cassatella}, A., {Altamore}, A., {Gonz{\'a}lez-Riestra}, R., {et~al.} 2000,
  Astron. Astrophys. Suppl. Ser., 141, 331

\bibitem[{{Cerruti-Sola} {et~al.}(1992){Cerruti-Sola}, {Cheng}, \&
  {Pallavicini}}]{1992A&A...256..185C}
{Cerruti-Sola}, M., {Cheng}, C.-C., \& {Pallavicini}, R. 1992, \aap, 256, 185

\bibitem[{{Cincunegui} {et~al.}(2007){Cincunegui}, {D{\'{\i}}az}, \&
  {Mauas}}]{2007astro.ph..3511C}
{Cincunegui}, C., {D{\'{\i}}az}, R.~F., \& {Mauas}, P.~J.~D. 2007, \aap, 469,
  309

\bibitem[{{Donahue} {et~al.}(1996){Donahue}, {Saar}, \&
  {Baliunas}}]{1996ApJ...466..384D}
{Donahue}, R.~A., {Saar}, S.~H., \& {Baliunas}, S.~L. 1996, \apj, 466, 384

\bibitem[{{Dravins} {et~al.}(1993){Dravins}, {Linde}, {Fredga}, \&
  {Gahm}}]{1993ApJ...403..396D}
{Dravins}, D., {Linde}, P., {Fredga}, K., \& {Gahm}, G.~F. 1993, \apj, 403, 396

\bibitem[{{Dravins} {et~al.}(1998){Dravins}, {Lindegren}, \&
  {Vandenberg}}]{1998A&A...330.1077D}
{Dravins}, D., {Lindegren}, L., \& {Vandenberg}, D.~A. 1998, \aap, 330, 1077

\bibitem[{{Fernandes} {et~al.}(1998){Fernandes}, {Lebreton}, {Baglin}, \&
  {Morel}}]{1998A&A...338..455F}
{Fernandes}, J., {Lebreton}, Y., {Baglin}, A., \& {Morel}, P. 1998, \aap, 338,
  455

\bibitem[{{Flower}(1996)}]{1996ApJ...469..355F}
{Flower}, P.~J. 1996, \apj, 469, 355

\bibitem[{{Frodesen} {et~al.}(1979){Frodesen}, {Skjeggestad}, \&
  {Tofte}}]{Frod79}
{Frodesen}, G.~A., {Skjeggestad}, O., \& {Tofte}, H. 1979, {Probability and
  Statistics in Particle Physics} (Universitetsforlaget)

\bibitem[{{Garhart} {et~al.}(1997){Garhart}, {Smith}, {Turnrose}, {Levay}, \&
  {Thompson}}]{1997IUENN..57....1G}
{Garhart}, M.~P., {Smith}, M.~A., {Turnrose}, B.~E., {Levay}, K.~L., \&
  {Thompson}, R.~W. 1997, IUE NASA Newsletter, 57, 1

\bibitem[{{Gray} \& {Baliunas}(1994)}]{1994ApJ...427.1042G}
{Gray}, D.~F. \& {Baliunas}, S.~L. 1994, \apj, 427, 1042

\bibitem[{{Gray} \& {Baliunas}(1995)}]{1995ApJ...441..436G}
{Gray}, D.~F. \& {Baliunas}, S.~L. 1995, \apj, 441, 436

\bibitem[{{Guedel} {et~al.}(1997){Guedel}, {Guinan}, \&
  {Skinner}}]{1997ApJ...483..947G}
{Guedel}, M., {Guinan}, E.~F., \& {Skinner}, S.~L. 1997, \apj, 483, 947

\bibitem[{{Guenther} \& {Demarque}(2000)}]{2000ApJ...531..503G}
{Guenther}, D.~B. \& {Demarque}, P. 2000, \apj, 531, 503

\bibitem[{{Hall} {et~al.}(2007){Hall}, {Lockwood}, \&
  {Skiff}}]{2007AJ....133..862H}
{Hall}, J.~C., {Lockwood}, G.~W., \& {Skiff}, B.~A. 2007, \aj, 133, 862

\bibitem[{{Hallam} {et~al.}(1991){Hallam}, {Altner}, \&
  {Endal}}]{1991ApJ...372..610H}
{Hallam}, K.~L., {Altner}, B., \& {Endal}, A.~S. 1991, \apj, 372, 610

\bibitem[{{Heath} \& {Schlesinger}(1986)}]{1986JGR....91.8672H}
{Heath}, D.~F. \& {Schlesinger}, B.~M. 1986, \jgr, 91, 8672

\bibitem[{{Henry} {et~al.}(1996){Henry}, {Soderblom}, {Donahue}, \&
  {Baliunas}}]{1996AJ....111..439H}
{Henry}, T.~J., {Soderblom}, D.~R., {Donahue}, R.~A., \& {Baliunas}, S.~L.
  1996, \aj, 111, 439

\bibitem[{{Hoeg} {et~al.}(1997){Hoeg}, {B{\"a}ssgen}, {Bastian}, {Egret},
  {Fabricius}, {Gro{\ss}mann}, {Halbwachs}, {Makarov}, {Perryman},
  {Schwekendiek}, {Wagner}, \& {Wicenec}}]{1997A&A...323L..57H}
{Hoeg}, E., {B{\"a}ssgen}, G., {Bastian}, U., {et~al.} 1997, \aap, 323, L57

\bibitem[{{Horne} \& {Baliunas}(1986)}]{1986ApJ...302..757H}
{Horne}, J.~H. \& {Baliunas}, S.~L. 1986, \apj, 302, 757

\bibitem[{{Jay} {et~al.}(1997){Jay}, {Guinan}, {Morgan}, {Messina}, \&
  {Jassour}}]{1997AAS...18912004J}
{Jay}, J.~E., {Guinan}, E.~F., {Morgan}, N.~D., {Messina}, S., \& {Jassour}, D.
  1997, in Bulletin of the American Astronomical Society, Vol.~29, Bulletin of
  the American Astronomical Society, 730

\bibitem[{{Judge} {et~al.}(2004){Judge}, {Saar}, {Carlsson}, \&
  {Ayres}}]{2004ApJ...609..392J}
{Judge}, P.~G., {Saar}, S.~H., {Carlsson}, M., \& {Ayres}, T.~R. 2004, \apj,
  609, 392

\bibitem[{{Kamper} \& {Wesselink}(1978)}]{1978AJ.....83.1653K}
{Kamper}, K.~W. \& {Wesselink}, A.~J. 1978, \aj, 83, 1653

\bibitem[{{Lachaume} {et~al.}(1999){Lachaume}, {Dominik}, {Lanz}, \&
  {Habing}}]{1999A&A...348..897L}
{Lachaume}, R., {Dominik}, C., {Lanz}, T., \& {Habing}, H.~J. 1999, \aap, 348,
  897

\bibitem[{{Messina} \& {Guinan}(2002)}]{2002A&A...393..225M}
{Messina}, S. \& {Guinan}, E.~F. 2002, \aap, 393, 225

\bibitem[{{Metcalfe} {et~al.}(2007){Metcalfe}, {Dziembowski}, {Judge}, \&
  {Snow}}]{2007MNRAS.tmpL..50M}
{Metcalfe}, T.~S., {Dziembowski}, W.~A., {Judge}, P.~G., \& {Snow}, M. 2007,
  \mnras, L50+

\bibitem[{{Nordstr{\"o}m} {et~al.}(2004){Nordstr{\"o}m}, {Mayor}, {Andersen},
  {Holmberg}, {Pont}, {J{\o}rgensen}, {Olsen}, {Udry}, \&
  {Mowlavi}}]{2004A&A...418..989N}
{Nordstr{\"o}m}, B., {Mayor}, M., {Andersen}, J., {et~al.} 2004, \aap, 418, 989

\bibitem[{{Oranje} \& {Zwaan}(1985)}]{1985A&A...147..265O}
{Oranje}, B.~J. \& {Zwaan}, C. 1985, \aap, 147, 265

\bibitem[{{Oranje} {et~al.}(1982){Oranje}, {Zwaan}, \&
  {Middelkoop}}]{1982A&A...110...30O}
{Oranje}, B.~J., {Zwaan}, C., \& {Middelkoop}, F. 1982, \aap, 110, 30

\bibitem[{{Perryman} {et~al.}(1997){Perryman}, {Lindegren}, {Kovalevsky},
  {Hoeg}, {Bastian}, {Bernacca}, {Cr{\'e}z{\'e}}, {Donati}, {Grenon}, {van
  Leeuwen}, {van der Marel}, {Mignard}, {Murray}, {Le Poole}, {Schrijver},
  {Turon}, {Arenou}, {Froeschl{\'e}}, \& {Petersen}}]{1997A&A...323L..49P}
{Perryman}, M.~A.~C., {Lindegren}, L., {Kovalevsky}, J., {et~al.} 1997, \aap,
  323, L49

\bibitem[{{Press} {et~al.}(1992){Press}, {Teukolsky}, {Vetterling}, \&
  {Flannery}}]{1992nrfa.book.....P}
{Press}, W.~H., {Teukolsky}, S.~A., {Vetterling}, W.~T., \& {Flannery}, B.~P.
  1992, {Numerical recipes in FORTRAN. The art of scientific computing}
  (Cambridge: University Press, |c1992, 2nd ed.)

\bibitem[{{Robrade} {et~al.}(2005){Robrade}, {Schmitt}, \&
  {Favata}}]{2005A&A...442..315R}
{Robrade}, J., {Schmitt}, J.~H.~M.~M., \& {Favata}, F. 2005, \aap, 442, 315

\bibitem[{{Robrade} {et~al.}(2007){Robrade}, {Schmitt}, \&
  {Hempelmann}}]{2007MmSAI..78..311R}
{Robrade}, J., {Schmitt}, J.~H.~M.~M., \& {Hempelmann}, A. 2007, Memorie della
  Societa Astronomica Italiana, 78, 311

\bibitem[{{Rutten}(1984)}]{1984A&A...130..353R}
{Rutten}, R.~G.~M. 1984, \aap, 130, 353

\bibitem[{{Rutten} {et~al.}(1991){Rutten}, {Schrijver}, {Lemmens}, \&
  {Zwaan}}]{1991A&A...252..203R}
{Rutten}, R.~G.~M., {Schrijver}, C.~J., {Lemmens}, A.~F.~P., \& {Zwaan}, C.
  1991, \aap, 252, 203

\bibitem[{{Saar} \& {Osten}(1997)}]{1997MNRAS.284..803S}
{Saar}, S.~H. \& {Osten}, R.~A. 1997, \mnras, 284, 803

\bibitem[{{Saffe} {et~al.}(2005){Saffe}, {G{\'o}mez}, \&
  {Chavero}}]{2005A&A...443..609S}
{Saffe}, C., {G{\'o}mez}, M., \& {Chavero}, C. 2005, \aap, 443, 609

\bibitem[{{Scargle}(1982)}]{1982ApJ...263..835S}
{Scargle}, J.~D. 1982, \apj, 263, 835

\bibitem[{{Schmitt} \& {Liefke}(2004)}]{2004A&A...417..651S}
{Schmitt}, J.~H.~M.~M. \& {Liefke}, C. 2004, \aap, 417, 651

\bibitem[{{Schrijver}(1987)}]{1987A&A...172..111S}
{Schrijver}, C.~J. 1987, \aap, 172, 111

\bibitem[{{Schrijver} {et~al.}(1989){Schrijver}, {Dobson}, \&
  {Radick}}]{1989ApJ...341.1035S}
{Schrijver}, C.~J., {Dobson}, A.~K., \& {Radick}, R.~R. 1989, \apj, 341, 1035

\bibitem[{{Schrijver} {et~al.}(1992){Schrijver}, {Dobson}, \&
  {Radick}}]{1992A&A...258..432S}
{Schrijver}, C.~J., {Dobson}, A.~K., \& {Radick}, R.~R. 1992, \aap, 258, 432

\end{thebibliography}
\end{document}